\documentclass[aps,prd,preprint,groupedaddress,showpacs]{revtex4-1}
\usepackage{amssymb,amsmath}
\usepackage[dvips]{graphicx}

\begin{document}

\title{Nonperturbative infrared effects for light scalar fields in de Sitter space}

\author{Takashi Arai}
\email[]{araitks@post.kek.jp}
\affiliation{KEK Theory Center, Tsukuba, Ibaraki 305-0801, Japan}
\affiliation{Graduate School of Mathematics, Nagoya University, Nagoya 464-8602, Japan}

\begin{abstract}
We study the $\phi^4$ scalar field theory in de Sitter space using the 2PI effective action formalism. This formalism enables us to investigate the nonperturbative quantum effects. We use the mean field and gap equations to calculate the physical mass and the effective potential. We find that nonperturbative infrared effects in de Sitter space produce a curvature-induced mass and work to restore the broken $Z_2$ symmetry.
\end{abstract}


\maketitle

\section{Introduction}
The study of quantum fields in de Sitter space has a long history. This is partly due to the high symmetry of de Sitter geometry. In fact, this symmetry enables us to exactly solve the theory regardless of its curved spacetime nature~\cite{Candelas}. Recently, quantum field theory in de Sitter space has again attracted attention due to measurements of the Cosmic Microwave Background radiation anisotropy and the discovery of the accelerating expansion of the present universe. It is well known that in de Sitter space, a propagator of massless scalar fields which minimally couple to the background geometry diverges due to infrared effects~\cite{Allen1,Allen2}. In other words, there is no well-defined propagator for a massless, minimally coupled scalar field in de Sitter symmetric states. This means that a perturbative calculation will break down which causes difficulties in computing rigorous quantum corrections for the inflationary universe. At present, whether this infrared divergence is physical entity or not is unexplained.

In an attempt to regulate the infrared divergence, a low-momentum cutoff could be introduced. This cutoff prescription corresponds to considering a local de Sitter geometry. This idea is justified because in a realistic inflationary scenario, an inflationary expansion phase ceases at a finite time. This cutoff partially breaks the de Sitter symmetry, and the propagator gets the logarithmic term of $a(\eta)$, a scale factor where $\eta$ is the conformal time. This fact means that physical quantities such as the energy-momentum tensor may become time dependent~\cite{Woodard1,Woodard2,Kitamoto}.

Another method is to investigate the effects of a resummation of infrared divergences. One example is an effective stochastic approach devised by Starobinsky and Yokoyama~\cite{Starobinsky}. This approach treats the field fluctuations of low-frequency modes which exceed the Hubble scale stochastically, and enables us to investigate the resummed effects of infrared terms. However the stochastic approach includes only the leading infrared terms at each order of perturbative expansion. On the other hand, the two-particle irreducible (2PI) formalism is adopted for a resummation of the $O(N)$ linear sigma model to estimate the effects of quantum corrections beyond one-loop level to the non-Gaussianity~\cite{Riotto}. In their paper, the large-$N$ limit is taken to simplify the equations, and again the stochastic assumption is used for solving the equations. Their research suggests that a self-interacting scalar field acquires an effective mass which prevents infrared divergences. More recently, the same theory is investigated by the $1/N$ expansion without the 2PI formalism or the stochastic assumption~\cite{Serreau}. This expansion resums the infinite series of so-called cactus loop diagrams. Again, it was shown that the ultralong-wavelength fluctuations of scalar fields generate a strictly positive curvature-induced mass, and forbid the existence of spontaneously symmetry broken states.

In this paper, we consider a resummation approach without the $1/N$ expansion from the basic principles of quantum field theory. The reason we avoid the $1/N$ expansion is as follows: this expansion is used for the investigation of the nonperturbative nature of quantum field theories. Using this method, it has been shown that for $O(N)$ field theories, spontaneous breaking of the global $O(N)$ symmetry is impossible, not only in flat space~\cite{Coleman,Abbott} but also in curved space~\cite{Anderson}. Hence, it is possible that effects such as dynamical mass generation and the absence of spontaneously broken states are not genuine effects of the models or the nature of curved geometry, but are simply due to the nature of the $1/N$ expansion. Thus, to separate the characteristic properties of the $1/N$ expansion, we should resum the loop diagrams without the $1/N$ technique. We will employ the 2PI resummation technique instead of the $1/N$ technique. For our purpose, multi-fields are unnecessary. For simplicity, we will thus study the infrared effects of the $\phi^4$ scalar field theory.

This paper is organized as follows. In Sec. II, we briefly review the infrared problem in de Sitter space. In Sec. III, we show how the 2PI formalism circumvents the infrared divergences. In Sec. IV, we derive the mean field and gap equations varying the 2PI effective action which is truncated up to the double bubble diagram with respect to the mean field and the full propagator. From these equations, we can identify the physical mass and calculate the effective potential. Sec. V is devoted to the discussion and conclusion. In this paper, we adopt the unit system of $c=\hbar=1$.

\section{Free field propagator in de Sitter space}
In this section, we briefly review the infrared problem for a massless, minimally coupled free field in de Sitter space. De Sitter space is represented by the following line element in terms of comoving spatial coordinates {\bf x} and conformal time $-\infty < \eta < 0$
\begin{equation}
ds^2=a(\eta)^2(d\eta^2-d\mathbf{x}^2),
\end{equation}
where $a(\eta)=-1/H \eta$ is a scale factor, and $H$ is a Hubble parameter constant.

Let us consider a free scalar field theory on this geometry with the action
\begin{equation}
S_{\mathrm{free}}=-\frac{1}{2}\int d^dx \sqrt{-g}\phi(x)(\Box+m^2+\xi R) \phi(x),
\end{equation}
where $R=d(d-1) H^2$ is the Ricci scalar curvature and $\xi$ is a conformal factor. 
Then, the quantum field $\phi(x)$ is expressed as the mode expansion of the comoving-momentum~\cite{Bunch}:
\begin{equation}
\phi(x)=\int \frac{d^3k}{(2\pi)^{3/2}} H\eta^{3/2}\frac{\sqrt{\pi}}{2}
\Bigl(a_{\mathbf{k}} H_{\nu}^{(1)}(k|\eta|)e^{i\mathbf{k}\cdot\mathbf{x}}
+a_{\mathbf{k}}^{\dagger} H_{\nu}^{(2)}(k|\eta|)e^{-i\mathbf{k}\cdot\mathbf{x}} \Bigr),
\end{equation}
where $\mathbf{k}$ is a comoving spatial momentum, $k=|\mathbf{k}|$, $H_{\nu}^{(1)}(z)$ is the Hankel function and $\nu=\bigl \{ [ (d-1)/2 ]^2-(m^2+\xi R)/H^2 \bigr\}^{1/2}$. The operators $a_{\mathbf{k}}$ and $a_{\mathbf{k}}^{\dagger}$ are the annihilation and creation operators, respectively, with $a_{\mathbf{k}}|0\rangle=0$ and $[a_{\mathbf{k}},a_{\mathbf{k'}}^{\dagger}]=\delta^{(3)}(\mathbf{k}-\mathbf{k'})$. For a massless, minimally coupled field, $\nu=3/2$, the Hankel function is expressed as an elementary function, $H_{3/2}^{(1)}(z)=-\sqrt{2/\pi z}(1+i/z)e^{iz}$. Thus, the two-point function is expressed as follows
\begin{equation}
\langle0|\phi(x)\phi(x')|0\rangle=\int \frac{d^3k}{(2\pi)^3} \frac{H^2 \eta \eta'}{2k}
\bigl(1-i\frac{1}{k\eta} \bigr)\bigl(1+i\frac{1}{k\eta'}\bigr)e^{-ik(\eta-\eta')+i\mathbf{k}\cdot(\mathbf{x}-\mathbf{x'})}.
\end{equation}
This expression is obviously infrared divergent. This is the infrared divergence for a massless, minimally coupled scalar field. This divergence is usually regulated in the following two ways discussed below.

The first way is to introduce a low-momentum cutoff $\Lambda$. This prescription corresponds to considering a local de Sitter geometry. In this case, the two-point function of a field with a small mass is expressed by~\cite{Allen1,Allen2}
\begin{equation}
G_0 (x,x')=\frac{H^2}{4\pi^2}\biggl[-\frac{1}{y}-\frac{1}{2} \log(-y)+\frac{1}{2}\log(a(\eta)a(\eta'))-\frac{1}{4}+\log2 +\mathcal{O}\Bigl(\frac{m^2}{H^2}\Bigr) \biggr],
\end{equation}
where we take $\xi=0$, $y(x,x')=\bigl[ (\eta-\eta')^2-|\mathbf{x}-\mathbf{x'}|^2\bigr]/\eta \eta'$ is a de Sitter invariant length and $\Lambda$ is taken to be $H$. This expression has the correct massless limit. The third term in this expression represents the de Sitter breaking term. This time-dependent term eventually breaks down the perturbative expansion.

The second method regulates the infrared divergence using a small-mass parameter. In this case, the de Sitter symmetry is retained, and the equation of the propagator is transformed to a differential equation of the de Sitter invariant length $y$:
\begin{equation}
a^4 H^2\Bigl[ y (4+y) \frac{d^2}{dy^2}+d(2+y) \frac{d}{dy}+\frac{m^2+\xi R}{H^2} \Bigr] G_0(x,x')=-i \delta(x-x').
\end{equation}
This equation is solved by the hypergeometric function~\cite{Candelas}
\begin{equation}
G_0(x,x')=\frac{H^{d-2}}{(4\pi)^{d/2}}\frac{\Gamma(\frac{d-1}{2}+\nu) \Gamma(\frac{d-1}{2}-\nu)}{\Gamma(\frac{d}{2})} {}_2\mathrm{F}_1\left[\tfrac{d-1}{2}+\nu,\tfrac{d-1}{2}-\nu,\tfrac{d}{2};1+\tfrac{y}{4}\right].
\end{equation}
If we take $\xi=0$, and expand the propagator around $m=0$, we obtain~\cite{Garbrecht}
\begin{equation}
G_0(x,x')=\frac{H^2}{4 \pi^2} \biggl[ -\frac{1}{y}-\frac{1}{2}\log (-y)+\frac{3 H^2}{2 m^2}-1+\log 2+\mathcal{O}\Bigl( \frac{m^2}{H^2}\Bigr) \biggr].
\end{equation}
Of course, this propagator does not have a de Sitter breaking term. If we take $m \rightarrow 0$, $G(x,x')$ will diverge due to infrared effects, i.e. the mass parameter works as a regularization parameter of the infrared divergence. In Sec. III, we will see how the 2PI formalism circumvents the infrared divergence.

\section{Two-particle irreducible formalism}
In this section, we review the two-particle irreducible formalism in a single-field scalar theory for the purpose of designating our notations and conventions, though this formalism is well known~\cite{Jackiw,Ramsey}. Note also that although we should adopt Schwinger-Keldysh formalism~\cite{Ramsey}, because of the nonequilibrium nature of de Sitter space, we drop the closed-time path indices to avoid notational complexity. This is justified since to our order of approximation, the Schwinger-Keldysh formalism is similar to the usual in-out formalism. That is, $G(x,x')$ in this paper coincides with both $G_{++}(x,x')$ and $G_{--}(x,x')$ in the Schwinger-Keldysh formalism~\cite{Ramsey,Riotto}.

\begin{figure}
\includegraphics[width=10cm,clip]{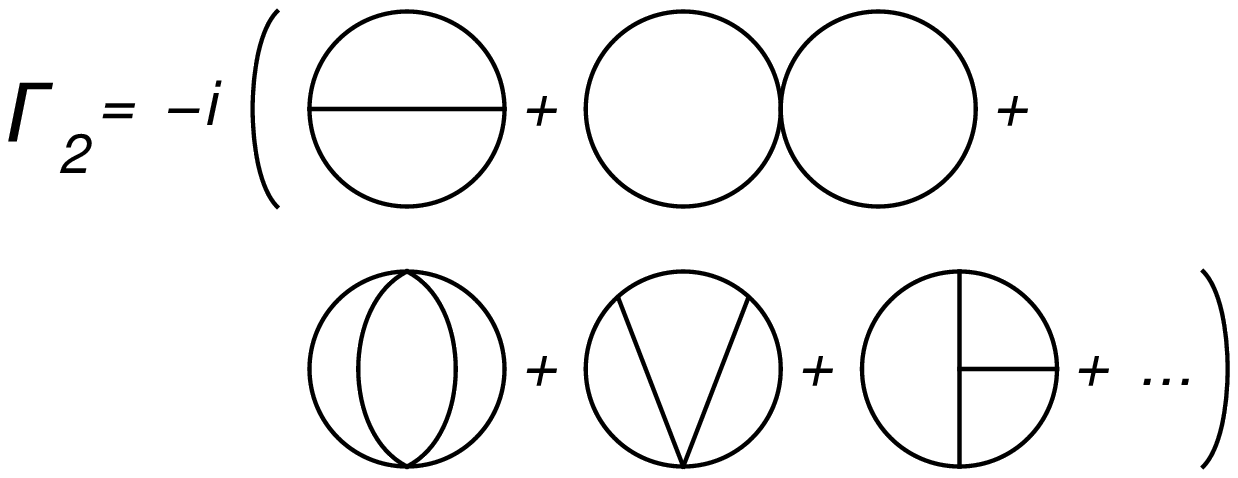}
\caption{\label{fig:2PI} Diagrammatic expansion for $\Gamma_2[\bar{\phi},G]$. These are the setting sun diagram, double bubble diagram and so on. Lines represent the propagator $G(x,x')$, and vertices terminating three lines are proportional to $\bar{\phi}$.}
\end{figure}

\begin{figure}
\includegraphics[width=3.9cm,clip]{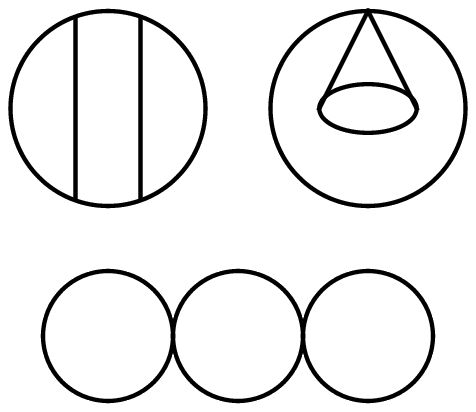}
\caption{\label{fig:1PI} Two-particle reducible graphs which do not contribute to $\Gamma_2[\bar{\phi},G]$ in a $\phi^4$ theory.}
\end{figure}

In the 2PI approach, we generalize the standard generating functional to include an additional double source term. Thus, we define the generating functional in the following way:
\begin{equation}
\begin{split}
Z[J,K]=\int D\phi \exp{} \Bigl( &iS[\phi]+i\int d^4x\sqrt{-g} J(x) \phi(x) \\
&+\frac{i}{2}\int d^4x \sqrt{-g} \int d^4x'\sqrt{-g'}\phi(x)K(x,x')\phi(x') \Bigr ).
\end{split}
\end{equation}
We also define $W[J,K]=-i \log Z[J,K]$. Then
\begin{equation}
\frac{1}{\sqrt{-g}}\frac{\delta W}{\delta J(x)}=\bar{\phi}(x),
\end{equation}
\begin{equation}
\frac{1}{\sqrt{-g}}\frac{1}{\sqrt{-g'}}\frac{\delta W}{\delta K(x,x')}=\frac{1}{2}\langle \phi(x)\phi(x')\rangle
                                                                                                             \equiv \frac{1}{2}(\bar{\phi}(x)\bar{\phi}(x')+G(x,x')),
\end{equation}
where $G(x,x')=\langle \varphi(x) \varphi(x') \rangle$, and $\varphi(x)=\phi(x)-\bar{\phi}$ is a shifted field. Now one can eliminate $J$ and $K$ in terms of $\bar{\phi}$ and $G$, and define the 2PI effective action as a double Legendre transformation of $W$.
\begin{equation}
\Gamma[\bar{\phi},G]=W[J,K]-\int d^4xJ(x)\frac{\delta W}{\delta J(x)}-\int d^4x \int d^4x' K(x,x') \frac{\delta W}{\delta K(x,x')}.
\end{equation}
This expression is rearranged to the following way~\cite{Jackiw,Ramsey}:
\begin{equation}
\Gamma[\bar{\phi},G]=S[\bar{\phi}]+\frac{i}{2}\log \mathrm{det}[G^{-1}]+\frac{i}{2}\int d^4x\sqrt{-g}\int d^4x'\sqrt{-g'}G_0^{-1}[\bar{\phi}](x,x')G(x',x)+\Gamma_2[\bar{\phi},G],
\end{equation}
where
\begin{equation}
i G_0^{-1}[\bar{\phi}](x,x')=\frac{1}{\sqrt{-g}}\frac{\delta^2S[\bar{\phi}]}{\delta\phi(x)\delta\phi(x')}\frac{1}{\sqrt{-g'}},
\end{equation}
and $\Gamma_2[\bar{\phi},G]$ is expressed by $(-i)$ times all of the two-particle irreducible vacuum diagrams with propagator $G$ and vertices given by a shifted action $S_{\mathrm{int}}$, defined by
\begin{equation}
S_{\mathrm{int}}[\varphi]=\sum_{n=3}^{\infty}\frac{1}{n!} \left( \prod_{i=1}^{n} \int d^4x_i  \right) \frac{\delta^n S[\bar{\phi}]}{\delta \phi(x_1) \cdots \delta\phi(x_n)} \varphi(x_1) \cdots \varphi(x_n).
\end{equation}
A two-particle irreducible diagram is a diagram which can not be cut in two by cutting only two internal lines, otherwise it is two-particle reducible. The diagrammatic expansion of $\Gamma_2[\bar{\phi},G]$ for $\phi^4$ theory is shown in Fig. \ref{fig:2PI}. The two-particle reducible diagrams at the three-loop level are also shown in Fig. \ref{fig:1PI}. The cactus and the ladder diagrams, for example, are not included in the 2PI vacuum diagrams. Thus $\Gamma_2[\bar{\phi},G]$ captures the nonperturbative contribution of these diagrams. In this formalism, the mean field equation and the gap equation are given by
\begin{equation}
\frac{\delta \Gamma[\bar{\phi},G]}{\delta G(x,x')}
=-\sqrt{-g}\sqrt{-g'}G^{-1}(x,x')+\sqrt{-g}\sqrt{-g'}G_0^{-1}[\bar{\phi}](x,x')-2 i \frac{\delta \Gamma_2}{\delta G(x,x')}=0,
\end{equation}
\begin{equation}
\frac{\delta \Gamma[\bar{\phi},G]}{\delta \bar{\phi}(x)}
=\frac{\delta S[\bar{\phi}]}{\delta \bar{\phi}(x)}+\frac{i}{2}\int d^4y\sqrt{-g}\int d^4 y'\sqrt{-g'} \frac{\delta G_0^{-1}[\bar{\phi}](y,y')}{\delta \bar{\phi}(x)}G(y',y)=0.
\end{equation}
From these equations, we can solve $G=G[\bar{\phi}]$, and an ordinary one-particle irreducible (1PI) effective action is obtained by inserting $G[\bar{\phi}]$ into $\Gamma[\bar{\phi},G]$. Various approximations can be made by truncating the diagrammatic expansion for $\Gamma_2[\bar{\phi},G]$. In particular, the 2PI effective action contains the commonly used one-loop, Hartree-Fock, and large-$N$ (in the case of the $O(N)$ model) approximations~\cite{Ramsey}.

In the standard effective action formalism, the effective action is expanded using the 1PI vacuum diagrams with the propagator given by $G_0$. In contrast, in the 2PI formalism, the effective action is expanded using the 2PI vacuum diagrams with the propagator given by $G$. Thus, in the 2PI formalism, contributions of infinitely many two-particle reducible diagrams are resumed into the 2PI diagrams. Moreover, one does not use the free field propagator in the calculation, and one does not need to know whether the free field propagator is well-defined or not.

\section{The contribution of double bubble diagram}
For the purpose of investigating nonperturbative infrared effects, we study a minimally coupled $\phi^4$ scalar field theory with the following action
\begin{equation}
S=-\int d^4x\sqrt{-g} \Bigl[ \frac{1}{2}\phi(x)(\Box+m^2)\phi(x)+\frac{\lambda}{4!}\phi^4(x) \Bigr].
\end{equation}

Here we consider an approximation scheme by including the double bubble diagram in the 2PI vacuum diagrams shown in Fig. \ref{fig:2PI}. For $\phi^4$ theory, this approximation corresponds to the Hartree-Fock approximation. Then the mean field and the gap equations are given by
\begin{equation}
-\sqrt{-g}\Bigl( \Box+m^2+\frac{\lambda}{6}\bar{\phi}^2+\frac{\lambda}{2}G(x,x)\Bigr) \bar{\phi}(x)=0,
\end{equation}
\begin{equation}
\sqrt{-g}\Bigl(\Box+m^2+\frac{\lambda}{2}\bar{\phi}^2(x)+\frac{\lambda}{2}G(x,x)\Bigr)G(x,y)=-i \delta(x-y).
\end{equation}
Quite recently, these equations were also derived from a mean field analysis~\cite{Prokopec}. To solve these equations, we utilize the de Sitter symmetry. If we assume that the vacuum state is de Sitter invariant, then $\bar{\phi}=\mathrm{const}$ and $G(x,x')$ will depend only on the de Sitter invariant length $y(x,x')$, i.e. $G(x,x')=G(y)$. The two-point function at the same spacetime point $G(x,x)$ is then a constant $G(0)$. Thus, the full propagator obeys the same equation as the free field one, even though the interaction is taken into account.

Here we can naturally identify the physical mass $m_{\mathrm{ph}}$ in analogy with the free propagator as
\begin{equation}
m^2_{\mathrm{ph}}=m^2+\frac{\lambda}{2}\bar{\phi}^2+\frac{\lambda}{2}G(x,x).
\end{equation}
The gap equation is given by
\begin{equation}
\sqrt{-g}(\Box+m_{\mathrm{ph}}^2)G(x,y)=-i \delta(x-y).
\end{equation}

The above mean field and gap equations are apparently divergent. To solve these equations, we have to renormalize them. In nonperturbative approximation schemes, renormalization is a nontrivial task, though some renormalization prescriptions exist~\cite{Stevenson,Paz,Rischke}. Among them, we adopt the $\overline{\mathrm{MS}}$ scheme. The divergent term $G(x,x)$ is regularized by dimensional regularization introducing the scale $\lambda \rightarrow \mu^{4-d} \lambda$. If we assume that the renormalized mass square $m_{\mathrm{ph}}^2$ is small compared to $H^2$ and use the small mass expansion of the propagator (see Appendix), the renormalization scale is introduced to
\begin{equation}
\begin{split}
\lambda G(x,x)& \rightarrow \Bigl(1+\frac{\epsilon}{2}\log\mu^2+\mathcal{O}(\epsilon^2) \Bigr)G(x,x), \\
&=\frac{H^2}{16\pi^2} \Biggl \{ \biggl (2-\frac{m_{\mathrm{ph}}^2}{H^2}\biggr ) \frac{2}{\epsilon}-2 \biggl [\gamma+\log \Bigl( \frac{H^2}{4 \pi \mu^2}\Bigr)\biggr ]
+\frac{6H^2}{m_{\mathrm{ph}}^2}-\biggl (\frac{23}{3}-4 \gamma \biggr )
+\mathcal{O}(\epsilon,\tfrac{m_{\mathrm{ph}}^2}{H^2}) \Biggr \}, \\
&=\frac{H^2}{16\pi^2} \Biggl \{ \biggl (2-\frac{m_{\mathrm{ph}}^2}{H^2}\biggr ) \frac{2}{\epsilon}-2 \log \Bigl( \frac{H^2}{4 \pi \overline{\mu}^2}\Bigr)
+\frac{6H^2}{m_{\mathrm{ph}}^2}-\biggl (\frac{23}{3}-4 \gamma \biggr )
+\mathcal{O}(\epsilon,\tfrac{m_{\mathrm{ph}}^2}{H^2}) \Biggr \},
\end{split}
\end{equation}
where $\gamma$ is the Euler-Mascheroni constant and $\bar{\mu}^2=4\pi\mu^2 e^{-\gamma}$ is the renormalization scale of the $\overline{\mathrm{MS}}$ scheme. Now we assume that we can drop the ultraviolet divergent pole terms so that the mean field and the gap equations are renormalized to
\begin{equation}
-\sqrt{-g}\Bigl(m_{\mathrm{ph}}^2-\frac{\lambda}{3}\bar{\phi}^2\Bigr)\bar{\phi}=0,
\end{equation}
\begin{equation}
\sqrt{-g}\Bigl(\Box+m_{\mathrm{ph}}^2\Bigr)G(x,y)=-i \delta(x-y).
\end{equation}
Here, the physical mass is renormalized to
\begin{equation}
m_{\mathrm{ph}}^2=m^2-b+\frac{\lambda}{2}\bar{\phi}^2+
\frac{\lambda H^2}{16\pi^2}\biggl(\frac{3 H^2}{m_{\mathrm{ph}}^2}\biggr),
\end{equation}
where we discard the terms $\mathcal{O}(m_{\mathrm{ph}}^2/H^2)$ and $b=\lambda H^2 \bigl[23/6-2 \gamma+\log(H^2/\bar{\mu}^2) \bigr]/16\pi^2$. 

First, we solve $m_{\mathrm{ph}}^2$ as a functional of $\bar{\phi}$. The mass equation is solved to
\begin{equation}
m_{\mathrm{ph}}^2(\bar{\phi}^2)=\frac{1}{2}\left(m^2-b+\frac{\lambda}{2}\bar{\phi}^2+\sqrt{(m^2-b+\tfrac{\lambda}{2}\bar{\phi}^2)^2+\frac{3\lambda H^4}{4\pi^2}}\right).
\end{equation}
Here, we have dropped the solution of the negative value because it is unphysical: this solution leads to the instability of the system. Note that our small mass expansion is justified only in the limited range of the renormalized parameters, $0<\lambda \ll1$ and $m^2/H^2 \ll 1$. This result means that we cannot set $m_{\mathrm{ph}}^2$ to $0$, and its minimum value is $m_{\mathrm{ph}}^2=\sqrt{3\lambda}H^2/4\pi$. This value is in perfect agreement with the estimate by the Hartree-Fock approximation~\cite{Starobinsky}.

\begin{figure}
\includegraphics[width=10cm,clip]{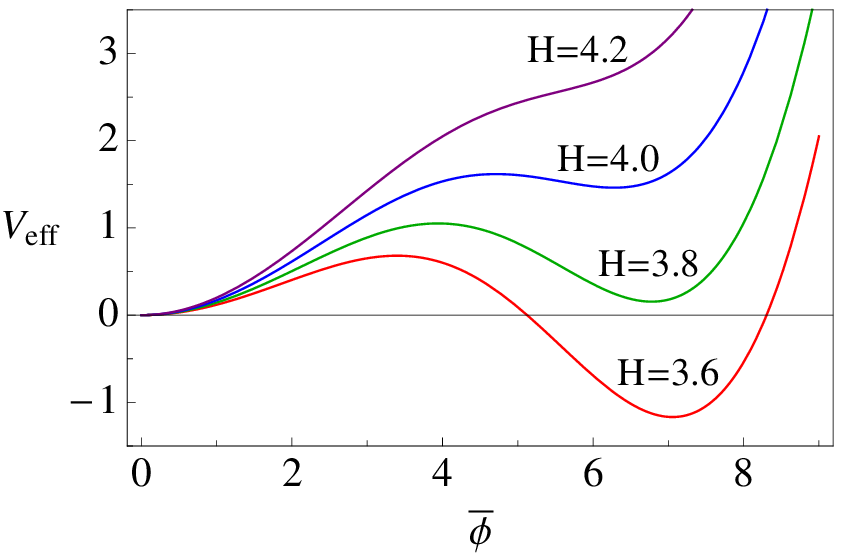}
\caption{\label{fig:EP} The effective potentials as a function of $\bar{\phi}$ for $\lambda=0.1$ and $\bar{\mu}=H$, all in the units of $|m|$ and $m^2<0$. The different lines show the potential with different values of $H$.}
\end{figure}

We now turn to the mean field equation to calculate the effective potential. Making use of the $m_{\mathrm{ph}}^2(\bar{\phi}^2)$ expression, we have
\begin{equation}
\Bigl(m_{\mathrm{ph}}^2(\bar{\phi}^2)-\frac{\lambda}{3}\bar{\phi}^2\Bigr)\bar{\phi}=0.
\end{equation}
This equation corresponds to $\partial V_{\mathrm{eff}}/\partial \bar{\phi}=\bigl[m_{\mathrm{ph}}^2(\bar{\phi}^2)-\lambda \bar{\phi}^2 /3 \bigr]\bar{\phi}=0$, and the condition for the minimum energy field configurations is $\bar{\phi}=0$, or $\bigl[m_{\mathrm{ph}}^2(\bar{\phi}^2)-\lambda \bar{\phi}^2/3\bigr]=0$. The latter is solved as

\begin{equation}
\bar{\phi}^2=\frac{3}{\lambda}\left[-(m^2-b)\pm \sqrt{(m^2-b)^2-\frac{3\lambda H^4}{8\pi^2}}\right].
\label{eq:determination}
\end{equation}
This solution is real and positive if $-m^2+b>0$ and $(-m^2+b)^2>3\lambda H^4/8 \pi^2$, that is, $-m^2+b>\sqrt{3\lambda H^4/8 \pi^2}$. This means that in de Sitter space, spontaneous symmetry breaking is possible~\cite{comment}, which is in contrast to the conclusion of the $1/N$ expansion~\cite{Serreau}. In fact, according to $\partial V_{\mathrm{eff}}/\partial \bar{\phi}=2\bar{\phi} \partial V_{\mathrm{eff}}/ \partial \bar{\phi}^2=0$, the de Sitter effective potential is calculated to be~\cite{Serreau}

\begin{equation}
V_{\mathrm{eff}}(\bar{\phi}^2)=\frac{1}{2}\int_0^{\bar{\phi}^2} dv \Bigl(m_{\mathrm{ph}}^2(v)-\frac{\lambda}{3}v\Bigr),
\end{equation}
where we omit an irrelevant constant term. This equation is integrated to

\begin{equation}
\begin{split}
V_{\mathrm{eff}}(\bar{\phi}^2)=\frac{(m^2-b)}{4}\bar{\phi}^2&-\frac{\lambda}{48}\bar{\phi}^4+\frac{1}{\lambda}\Biggl[
\frac{(m^2-b+\frac{\lambda}{2}\bar{\phi}^2)}{4}\sqrt{(m^2-b+\frac{\lambda}{2}\bar{\phi}^2)^2+\frac{3\lambda H^4}{4\pi^2}} \\
&-\frac{(m^2-b)}{4}\sqrt{(m^2-b)^2+\frac{3\lambda H^4}{4\pi^2}}
+\frac{3\lambda H^4}{16\pi^2} \log \biggl(\frac{m_{\mathrm{ph}}^2(\bar{\phi}^2)}{m_{\mathrm{ph}}^2(0)} \biggr) \Biggr].
\end{split}
\end{equation}
The behavior of the effective potential as a function of $\bar{\phi}$ near the phase transition is displayed in Fig. \ref{fig:EP}.
These graphs indicate that the (local) minima of the potential correspond to the positive branch in Eq. (\ref{eq:determination}). These minima may be a true or a metastable vacuum as shown in Fig. \ref{fig:EP}. These graphs also show that the effects of de Sitter geometry work to restore the broken $Z_2$ symmetry, and that the phase transition is of a first-order.

\section{Discussion and Conclusions}
In conclusion, we verify that nonperturbative infrared quantum effects generate a strictly positive curvature-induced mass for $\phi^4$ scalar field theory on full de Sitter geometry. In contrast to the $1/N$ analysis, spontaneous symmetry breaking is possible. Nonperturbative infrared effects work to restore its spontaneously broken potential, and the phase transition is of a first-order. In our analysis, we restrict our attention to the limited range of renormalized parameters in the $\overline{\mathrm{MS}}$ scheme, $0<\lambda \ll1$ and $m^2/H^2 \ll 1$. We stress that this range of validity is sufficient for an analysis of the light fields. We derive these results from the basic  principles of quantum field theory without the large-$N$ limit, and we do not use any assumptions as is done in the stochastic approach. These results are obtained using the Hartree-Fock approximation. However, it is well known that the Hartree approximation predicts a first-order phase transition at finite temperature, and further investigation beyond this approximation lead to a second-order phase transition~\cite{Verschelde,Reinosa}. It is important to investigate whether the first-order phase transition remains true beyond the Hartree-Fock approximation. This demands new approximation assumptions~\cite{Garbrecht}. 

From these results, we believe that the absence of spontaneous symmetry breaking is not the nature of de Sitter geometry, but is a property of the $1/N$ expansion. However, the physical mass is always positive whether the spontaneous symmetry breaking may occur or not. This is a result of the genuine geometrical nature of de Sitter space. One cannot take $m_{\mathrm{ph}}=0$ in this model, and infrared divergence in a de Sitter propagator is naturally self-regulated due to the dynamically generated curvature-induced square mass. That is, infrared divergences are fictitious, and artificial cutoff parameters are not necessary. They arise from our perturbative expansion around $m=0$. Further study of the nonperturbative infrared effects in other models is also important. It would be interesting to investigate whether the self-regulation mechanism takes place similar to the dynamical mass generation in other models.

\begin{acknowledgments}
I would like to thank H. Kanno for helpful discussions and comments. I also thank J. Serreau for useful comments. 
This work was supported by the Grant-in-Aid for Nagoya University Global COE Program, "Quest for Fundamental Principles in the Universe: from Particles to the Solar System and the Cosmos", from the Ministry
of Education, Culture, Sports, Science and Technology of Japan.
\end{acknowledgments}

\appendix*
\section{Small mass expansion of a coincident propagator}

In de Sitter space, a free propagator of a minimally coupled scalar field is expressed with the hypergeometric function~\cite{Candelas}

\begin{equation}
G(x,x')=\frac{H^{d-2}}{(4\pi)^{d/2}}\frac{\Gamma(\frac{d-1}{2}+\nu) \Gamma(\frac{d-1}{2}-\nu)}{\Gamma(\frac{d}{2})} {}_2\mathrm{F}_1\left[\tfrac{d-1}{2}+\nu,\tfrac{d-1}{2}-\nu,\tfrac{d}{2};1+\tfrac{y}{4}\right], 
\end{equation}
where $\nu=\bigl\{ [ (d-1)/2]^2-m^2/H^2 \bigr\}^{1/2}$. By considering the same spacetime point $y=0$, the formula of the hypergeometric function ${}_2 \mathrm{F}_1(a,b,c;1)=\Gamma(c)\Gamma(c-a-b)/[\Gamma(c-a)\Gamma(c-b)]$ leads to
\begin{equation}
G(x,x)=\frac{H^{d-2}}{(4\pi)^{d/2}} \Gamma(1-\tfrac{d}{2})\frac{\Gamma(\frac{d-1}{2}+\nu) \Gamma(\frac{d-1}{2}-\nu)}{\Gamma(\frac{1}{2}+\nu) \Gamma(\frac{1}{2}-\nu)}.
\end{equation}
Introducing a dimensional regularization parameter $\epsilon=4-d$, we expand $\nu$ in powers of $\epsilon$ and a small mass parameter $(m^2/H^2)$
\begin{equation}
\nu=\frac{3}{2}-s(\epsilon,\tfrac{m^2}{H^2}),
\end{equation}
\begin{equation}
s(\epsilon,\tfrac{m^2}{H^2})=\tilde{s}+s^{\dagger}(\epsilon)+\frac{1}{2}\epsilon, 
\end{equation}
\begin{equation}
\tilde{s}=\frac{3}{2}\sum_{n=1}^{\infty}\frac{1}{n!}\frac{2^n}{9^n}(2n-3)!!\biggl(\frac{m^2}{H^2}\biggr)^n, 
\end{equation}
\begin{equation}
s^{\dagger}(\epsilon)=\frac{1}{2}\sum_{n=1}^{\infty}\frac{1}{n!}\frac{2^n}{9^n}(2n-3)!!(2n-1)\biggl(\frac{m^2}{H^2}\biggr)^n \epsilon+\mathcal{O}(\epsilon^2).
\end{equation}
This expansion also enables us to expand $G(x,x)$ in powers of $\epsilon$ and $(m^2/H^2)$. First, we expand $G(x,x)$ in powers of the regularization parameter $\epsilon$,

\begin{equation}
\begin{split}
G(x,x)&=\frac{H^{d-2}}{(4\pi)^{d/2}} \Gamma(-1+\tfrac{\epsilon}{2})\frac{\Gamma(\tilde{s}+s^{\dagger}) \Gamma(3-\tilde{s}-s^{\dagger}-\epsilon)}{\Gamma(2-\tilde{s}-s^{\dagger}-\frac{\epsilon}{2}) \Gamma(-1+\tilde{s}+s^{\dagger}+\frac{\epsilon}{2})}, \\
           &=\frac{H^{d-2}}{(4\pi)^{d/2}} \biggl[-\frac{2}{\epsilon}-1+\gamma+\mathcal{O}(\epsilon)\biggr] \\
           &\hspace{0.5cm} \biggl [
           \frac{\Gamma(\tilde{s}) \bigl[1+\psi(\tilde{s})s^{\dagger} \bigr]
                    \Gamma(3-\tilde{s}) \bigl[1-\Gamma(3-\tilde{s}) (s^{\dagger}+\epsilon) \bigr]}
                    {\Gamma(2-\tilde{s}) \bigl[1-\psi(2-\tilde{s})(s^{\dagger}+\frac{\epsilon}{2}) \bigr]
                    \Gamma(\tilde{s}-1) \bigl[1+\psi(\tilde{s}-1)(s^{\dagger}+\frac{\epsilon}{2}) \bigr]}
                    +\mathcal{O}(\epsilon^2) \biggr ], \\
          &=\frac{H^2}{16\pi^2} \left[ S_1\frac{2}{\epsilon} -S_1 \log\Bigl(\frac{H^2}{4\pi}\Bigr)+\frac{2}{\tilde{s}}+2 S_2 S_+           +2S_3+S_1(1-\gamma)+\mathcal{O}(\epsilon) \right], 
\end{split}
\end{equation}
where
\begin{equation}
S_1=2-\frac{m^2}{H^2},
\end{equation}
\begin{equation}
S_2=2 \tilde{s}-3, 
\end{equation}
\begin{equation}
S_3=-\frac{9}{2}+\frac{5}{2}\tilde{s}-\frac{1}{2}\Bigl(2-\frac{m^2}{H^2} \Bigr)[\psi(1+\tilde{s})+\psi(1-\tilde{s})],
\end{equation}
\begin{equation}
S_+=\frac{1}{2}\sum_{n=1}^{\infty}\frac{1}{n!}\frac{2^n}{9^n}(2n-3)!!(2n-1)\biggl(\frac{m^2}{H^2}\biggr)^n, 
\end{equation}
$\gamma$ is the Euler-Mascheroni constant and $\psi(x)$ is the digamma function. Then, we expand this expression in powers of $(m^2/H^2)$. If we retain up to $\mathcal{O}(m^2/H^2)$ and $\mathcal{O}(\epsilon^0)$, then
\begin{equation}
\begin{split}
G(x,x)=\frac{H^2}{16\pi^2}\Biggl \{ &-\biggl (\frac{m^2}{H^2}-2\biggr ) \frac{2}{\epsilon}
+\biggl (\frac{m^2}{H^2}-2\biggr ) \biggl (\gamma+\log \Bigl( \frac{H^2}{4 \pi}\Bigr)\biggr ) \\
&+\frac{6H^2}{m^2}-\biggl (\frac{23}{3}-4 \gamma \biggr )-\biggl (\frac{2}{27}+2 \gamma \biggr ) \frac{m^2}{H^2}
+\mathcal{O}(\epsilon,\bigl(\tfrac{m^2}{H^2}\bigr)^2) \Biggr \}. 
\end{split}
\end{equation}

\bibliography{basename of .bib file}

\begin{thebibliography}{99}
\bibitem{comment}
  A different conclusion is reached in Ref.~\cite{Paz}, where infinitesimal bare parameters are adopted for renormalization. Moreover, the crucial difference is that the conformal factor must be finite, i.e. their model is not a minimally coupled field.
\end{thebibliography}

\end{document}